\documentclass[%
 aip,
 amsmath,amssymb,
 reprint,%
]{revtex4-2}

\usepackage{graphicx}
\usepackage{dcolumn}
\usepackage{bm}

\usepackage[utf8]{inputenc}
\usepackage[T1]{fontenc}
\usepackage{txfonts} 
\usepackage{mathptmx}
\usepackage{mathrsfs}
\usepackage{etoolbox}
\usepackage{physics}
\usepackage{lipsum}
\usepackage{isotope}
\usepackage{xcolor}

\usepackage{hyperref}
\hypersetup{
    colorlinks=true,
    linkcolor=blue,
    filecolor=magenta,      
    urlcolor=cyan,
}

\newcommand{\mnj}{m_{N_j}}

\newcommand{\E}{\hat{E}}
\renewcommand{\H}{\mathscr{H}}
\newcommand{\covd}{\nabla_{\mathbf{B}}}
\newcommand{\B}{\mathbf{B}}
\renewcommand{\tr}{{}^{\rm T}}

\AtBeginDocument{\mathcode`v=\varv} 
\begin{document}

\preprint{AIP/123-QED}

\title[]{Computation of NMR shieldings at the CASSCF level using gauge-including atomic orbitals and Cholesky decomposition}

\author{Tommaso Nottoli}%
 \email{tommaso.nottoli@phd.unipi.it}
\affiliation{Dipartimento di Chimica e Chimica Industriale, Universit\`{a} di Pisa, Via G. Moruzzi 13, I-56124 Pisa, Italy}

\author{Sophia Burger}
 \email{soburger@uni-mainz.de}
\affiliation{Department Chemie, Johannes Gutenberg-Universit{\"a}t Mainz, Duesbergweg 10-14, D-55128 Mainz, Germany}

\author{Stella Stopkowicz}%
 \email{stella.stopkowicz@uni-mainz.de}
\affiliation{Department Chemie, Johannes Gutenberg-Universit{\"a}t Mainz, Duesbergweg 10-14, D-55128 Mainz, Germany}
\affiliation{Fachrichtung Chemie, Universit{\"a}t des Saarlandes, Campus B2.2, D-66123 Saarbr{\"u}cken, Germany }

\author{J{\"u}rgen Gauss}%
 \email{gauss@uni-mainz.de}
\affiliation{Department Chemie, Johannes Gutenberg-Universit{\"a}t Mainz, Duesbergweg 10-14, D-55128 Mainz, Germany}

\author{Filippo Lipparini}%
 \email{filippo.lipparini@unipi.it}
\affiliation{Dipartimento di Chimica e Chimica Industriale, Universit\`{a} di Pisa, Via G. Moruzzi 13, I-56124 Pisa, Italy}

\date{\today}

\begin{abstract}
We present an implementation of coupled-perturbed complete active space self-consistent field (CP-CASSCF) theory for the computation of nuclear magnetic resonance chemical shifts using gauge-including atomic orbitals and Cholesky decomposed two-electron integrals. The CP-CASSCF equations are solved using a direct algorithm where the magnetic Hessian matrix-vector product is expressed in terms of one-index transformed quantities. 
Numerical tests on systems with up to about 1300 basis functions provide information regarding both the computational efficiency and limitations of our implementation. 
\end{abstract}

\maketitle

\section{\label{sec:intro}Introduction}
Ab initio calculations of nuclear magnetic resonance (NMR) chemical shifts\cite{nmr_book} represent a powerful tool for the interpretation of experimental NMR spectra and for elucidating chemical structures and conformations. It is well known that the accurate prediction of NMR shieldings requires to take into account dynamic correlation\cite{Gauss93,Gauss1995,gauss2003electron} most often through the use of the existing hierarchy of coupled-cluster methods.\cite{cc_book}
Nevertheless, even the use of such accurate techniques may give erroneous results for systems that possess a significant multi-reference character. An established starting point for the treatment of multi-reference -- also called strongly correlated -- systems is given by the complete active space--self-consistent field
(CASSCF) method.\cite{roos_taylor1980,Werner1987,Shepard1987,Roos1987} With an adequate choice for the active space, this approach is able to qualitatively describe transition-metal complexes, radicals, molecules with stretched bonds and other systems where single-reference approaches fail. 

During the eighties and nineties, multiconfigurational (MC) SCF linear-response theory was developed\cite{dalgaard1980td,yeager1979mcscf_resp} and applied to the study of excitation energies, transition moments, and other excited state properties.\cite{olsen1985linear_resp,Jorgensen1988} At the same time, analytical second derivative formulations for a general MCSCF wave function were developed\cite{jorgensen1983,camp1983,hoffmann1984} where perturbed molecular and configurational parameters are obtained by solving a set of linear equations called coupled-perturbed (CP). The solution of this linear system of equations was later implemented using a direct formulation\cite{Almloef1985,Yamamoto1996,Bernhardsson1999} where the response Hessian matrix was never explicitly calculated. Concerning the calculation of magnetic properties for MCSCF references through the use of perturbation-dependent basis functions, namely gauge-including atomic orbitals (GIAOs), also called London orbitals,\cite{london1937,hameka58,ditchfield72,Wolinski90} implementations have been reported for NMR chemical shifts, magnetizabilities and the related rotational g-tensor.\cite{Ruud1994,Ruud_mag1995,Ruud_mag1997} In particular, the first description of the GIAO-MCSCF theory for shieldings and its implementation exploiting direct techniques was presented by Ruud {\it et al.}\cite{Ruud1994} It should be also noted that a MCSCF scheme for the computation of shieldings with a different treatment of the gauge-origin problem, that is by using individual gauges for localized orbitals (IGLOs), was reported by van W\"{u}llen and Kutzelnigg\cite{mc_iglo} in the same years.

CASSCF calculations and the subsequent solution of the corresponding CP equations are hampered by the combinatorial scaling exhibited by the full-CI (FCI) problem in the active space. For this reason, standard direct FCI implementations as the one reported in the present paper can be routinely applied only to small/medium sized active spaces, that is, up to 14 electrons in 14 orbitals. Under these circumstances, the main bottleneck in the calculation is given by the storage and manipulation of the two-electron repulsion integral (ERI) matrix, {\it i.e.}, the operations needed for the orbital optimization. The computational cost required to solve the CP-CASSCF equations are similar to the one of a conventional second-order CASSCF optimization algorithm. In particular, ERIs with two external indices are needed and the computational scaling is $\mathcal{O}((N_{\rm int}+N_{\rm act})N^4_{\rm b})$, $N_{\rm int}$ and $N_{\rm act}$ being the number of inactive and active orbitals, respectively, and $N_{\rm b}$ the total number of basis functions. 

In the last few years special emphasis has been placed on the development and rewriting of efficient and fast quantum-chemistry codes that are able to describe larger molecular systems. As for what concerns CASSCF second-order properties, one of the first efforts in this direction was made by Dudley {\it et al.}\cite{Dudley2006} who reported the implementation of a parallelized scheme for the solution of the CP-MCSCF equations. More recently, Snyder {\it et al.}\cite{Snyder2015,snyder2015atomic,Snyder2017} convincingly presented an implementation that exploits the sparsity in the atomic orbital (AO) basis and parallelizes the calculation over graphical processing units (GPU). In their paper, they report an observed reduction of the computational scaling from $\mathcal{O}(N^4_{\rm b})$ to $\mathcal{O}(N^2_{\rm b})$, thus allowing calculations on systems with up to almost 7000 basis functions. Furthermore, Helmich-Paris\cite{HelmichParis2019} recently reported results of linear-response CASSCF calculations on systems with more than 2000 basis functions. Despite being a linear-response formulation, such work shares similarities with the present implementation since it implements a direct algorithm based on the computation of one-index transformed Fock matrices, as it will also be shown in our derivation. In order to reduce computational cost, the work of Helmich-Paris exploits the resolution of the identity\cite{Whitten1973,Vahtras1993,neese2003improvement} and the overlap-fitted chain-of-spheres (COSX) approximation.\cite{neese2009efficient,izsak2011overlap} 

In this contribution, we exploit the Cholesky decomposition (CD) of the two-electron integrals.\cite{Beebe1977,Koch2003} This technique possesses the remarkable property of giving approximate results that can be easily improved by lowering the decomposition threshold. The formal scaling of the AO to MO integral transformation is reduced and is no longer the bottleneck. Conversely, special care has to be given to the contractions required to build Fock matrices and intermediates. Most importantly, the memory requirement is highly reduced with respect to a standard calculation. The CD has been widely used to accelerate the computation of energies and gradients at various levels of theory,\cite{Bozkaya2014,Aquilante2008,bozkaya2016mp2,blaschke2021,bostrom2014,Delcey2014,Feng2019,schnack2022} but so far little has been done for the computation of second-order properties, especially magnetic ones. The first application of the CD for the calculation of NMR shieldings has been only recently reported by some of us within a GIAO based second-order Møller-Plesset perturbation theory (GIAO-MP2) treatment.\cite{Burger2021}

In this paper, we discuss the implementation of a CD-based GIAO-CASSCF theory for the calculation of NMR chemical shifts within the {\sc CFour} program package.\cite{Matthews2020,cfour} In section \ref{sec:theory}, we briefly review the main equations of GIAO-CASSCF
theory and we describe the implementation of CD for the solution of the CP-CASSCF equations focusing on the handling of perturbed Cholesky vectors. In section \ref{sec:numerical}, we report our results showing both the accuracy of CD for the computation of NMR shieldings and the efficiency of our implementation in treating medium-sized systems. Section \ref{sec:conclusion} presents some concluding remarks.  

\section{\label{sec:theory}Theory and Implementation}
In this section we briefly review GIAO-CASSCF theory outlining the main equations. Then the Cholesky decomposition of perturbed integrals is introduced and the implementation of CD-GIAO-CASSCF is discussed. The following conventions for the indices are used in the following:
\begin{itemize}
    \item $\mu, \nu, \rho, \dots$ refer to AOs;
    \item $p, q, r, \dots$ refer to generic MOs;
    \item $i, j, k, \dots$ refer to inactive MOs;
    \item $u, v, x, \dots$ refer to active MOs;
    \item $a, b, c, \dots$ refer to external MOs;
    \item $I, J, \dots$ refer to either state functions or Slater determinants;
    \item $P, Q, \dots$ refer to the Cholesky vectors.
\end{itemize}

\subsection{GIAO-CASSCF theory}
The CASSCF wave function can be defined using the exponential unitary ansatz
\begin{equation}
    \ket{\Psi} = e^{i\hat{\kappa}}e^{i\hat{S}}\ket{0},
\end{equation}
where $\ket{0}$ is the multideterminantal reference state, $i$ the imaginary unit, and $\hat{\kappa}$ and $\hat{S}$ are the Hermitian orbital and configurational rotation operators defined as
\begin{align}
    \hat{\kappa} = \sum_{p>q}\kappa_{pq}\hat{E}^{+}_{pq},\\
    \hat{S} = \sum_{J\neq0}S_J(\ket{J}\bra{0} + \ket{0}\bra{J}).
\end{align}
Here, $\hat{E}^{+}_{pq} = \hat{E}_{pq} + \hat{E}_{qp}$ is the symmetric singlet excitation operator, while $\ket{J}$ represents a generic state orthogonal to the reference one. This functional form is more general than the one used in Ref. \citenum{Lipparini2016} and \citenum{Nottoli2021}, as we aim at describing the effect of a magnetic perturbation that introduces imaginary terms into the Hamiltonian.\cite{helgaker1991} Furthermore, to remove the gauge-origin dependence that affects the calculation of magnetic properties with finite basis sets, we resort as usual to the use of GIAOs.\cite{london1937,hameka58,ditchfield72,Wolinski90} This means that the basis set explicitly depends on the perturbation, {\it i.e.}, the external magnetic field ($\B$). To treat the magnetic-field dependence of the AO basis set, we follow the formalism proposed by Helgaker and Jørgensen\cite{Helgaker1984} that transfers the AO dependence on the perturbation to the MO overlap matrix. In other words, we require the MOs to stay orthonormal for any value of the perturbing field. It can be shown that this is achieved by introducing a modified gradient that enforces orthonormality of the MOs: 
\begin{equation}\label{eq:cov_der}
    \nabla_{x} h = \pdv{h}{x} - \frac{1}{2}\left\{\pdv{\ln{S}}{x}, h\right\}.
\end{equation}
In Eq.~(\ref{eq:cov_der}), curly braces are a compact representation for a one-index transformation, {\it e.g.}
\begin{equation}
    \{A,B\}_{pq} = \sum_r \left(A_{pr}B_{rq} + A^*_{qr}B_{pr}\right).
\end{equation}
In the definition of the modified gradient in Eq.~(\ref{eq:cov_der}), we adopt the so-called symmetric connection. An extended discussion on the topic of orbital connections can be found in the literature.\cite{Simons1984,Helgaker1984,Olsen1995,Ruud1995}

All terms that stem from the second part of Eq.~(\ref{eq:cov_der}) can be viewed as originating from the reorthogonalization of the MOs. As usual, differentiated quantities are then evaluated in the limit of vanishing perturbation. In this limit, the derivative of the logarithm of the overlap matrix can be shown to be equal to the derivative of the overlap matrix itself. We also note that Eq.~(\ref{eq:cov_der}) becomes the usual gradient when the overlap matrix does not depend on the perturbation $x$. 
The NMR chemical shieldings can be defined via the analytical second derivative of the variational CASSCF energy, $\mathcal{E}$, with respect to the components of an external magnetic field and the nuclear magnetic dipole moment:\cite{Ruud1994}
\begin{equation}\label{eq:chsh}
    \sigma_{ji} =  \nabla_{B_i}\pdv{\mathcal{E}}{\mnj} + \sum_{p>q}\pdv{\mathcal{E}}{\mnj}{\kappa_{pq}}\dv{\kappa_{pq}}{B_i} + \sum_{J\neq0} \pdv{\mathcal{E}}{\mnj}{S_{J}}\dv{S_{J}}{B_i}.
\end{equation}
In Eq.~(\ref{eq:chsh}), the first term is a static contribution, while the other two terms stem from chain rule differentiation and include the wave function response to the perturbation and will be called dynamic terms. 
Given the fact that the GIAOs do not depend on the magnetic dipole moments, it is convenient to perform first the differentiation with respect to $\mnj$ and then with respect to $B_i$ such that the static contribution takes the following form
\begin{equation}\label{eq:static}
    \nabla_{\B}\pdv{\mathcal{E}}{\mathbf{m}} = \sum_{pq}\left[h_{pq}^{(\B,\mathbf{m})} -\frac{1}{2}\sum_{r}(S^{(\B)}_{pr}h^{(\mathbf{m})}_{rq} + {S^{(\B)}}^*_{qr}h^{(\mathbf{m})}_{pr})\right]\gamma_{pq},
\end{equation}
where the superscript between parentheses is a shorthand notation for the explicit derivative of the integrals in the GIAO basis, {\it e.g.}
\begin{equation}
    h_{pq}^{(\B,\mathbf{m})} = \sum_{\mu\nu}c^*_{\mu p}\pdv{h_{\mu\nu}}{\B}{\mathbf{m}}c_{\nu q}
\end{equation}
with $c_{\mu p}$ as the MO coefficients.
Explicit expressions for the differentiated one-electron GIAO integrals are given by
\begin{align*}
    & \pdv{h_{\mu\nu}}{m_{N_j}} = \frac{1}{c}\bra{\chi_\mu}\frac{L_{N_j}}{|\mathbf{r}-\mathbf{R}_N|^3}\ket{\chi_\nu},\\
    & \pdv{h_{\mu\nu}}{B_i}{m_{N_j}} = \frac{1}{2c^2}\bra{\chi_\mu}\frac{\mathbf{r}\cdot(\mathbf{r}-\mathbf{R}_N)\delta_{ij} - \mathbf{r}_i(\mathbf{r}-\mathbf{R}_N)}{|\mathbf{r}-\mathbf{R}_N|^3}\ket{\chi_\nu},
\end{align*}
$L_{N_j}$ being the $j$-th component of the angular-momentum operator around nucleus $N$, $\mathbf{R}_N$ the position vector of nucleus $N$, and $c$ the speed of light in atomic units.
We use bold symbols to refer to the whole vector or array and implicitly assume that all derivatives are evaluated at zero field.

The calculation of the response terms of Eq.~(\ref{eq:chsh}) requires the solution of the coupled-perturbed CASSCF (CP-CASSCF) equations
\begin{equation}\label{eq:cpcas}
    \mathbf{G}\dv{\bm{\lambda}}{\B} = -\nabla_{\B}\pdv{\mathcal{E}}{\bm{\lambda}},
\end{equation}
where $\bm{\lambda}$ is a joint parameter including both $\kappa_{pq}$ and $S_J$ and where $\mathbf{G}=\pdv{\mathcal{E}}{\bm{\lambda}}{\bm{\lambda}}$ is the magnetic CASSCF Hessian.

The CP-CASSCF equations are solved in a direct way without the explicit computation and storage of the magnetic CASSCF Hessian. For NMR computations we use a plain preconditioned conjugate gradient (PCG) algorithm being the mathematically optimal choice for the solution of a symmetric, positive-definite linear system of equations. The diagonal Hessian is used as preconditioner. 

The right hand side of Eq.~(\ref{eq:cpcas}) is the derivative of the CASSCF gradient ($\mathbf{g}=\pdv{\mathcal{E}}{\bm{\lambda}}$) with respect to the external magnetic field, which can be written as follows:
\begin{align}
    & \covd g_{pq} = 2i(\covd F_{pq} + \covd F_{qp}), \label{eq:grad_mo}\\
    & \covd g_I =2i\bra{\Phi_I}\sum_{uv}\E_{uv} \covd F^{I}_{uv}+\frac{1}{2}\sum_{uvxy}\hat{e}_{uvxy}\covd g_{uvxy}\ket{0} \label{eq:grad_ci}.
\end{align}
Eq.~(\ref{eq:grad_mo}) represents the molecular part of the gradient while Eq.~(\ref{eq:grad_ci}) represents the configurational one, with $\ket{\Phi_I}$ being a generic Slater determinant. $g_{uvxy}$ is the active part of two-electron integral matrix written using Mulliken's convention. $F_{pq}$ is the generalized Fock matrix that can be defined in terms of the inactive and active Fock matrices, $F^I_{pq}$ and $F^A_{pq}$ respectively, and of the $Q$ matrix. The derivatives of these three intermediates with respect to $\B$, using the rule specified by Eq.~(\ref{eq:cov_der}), are given by
\begin{eqnarray}\label{eq:fi_b}
    \covd F^I_{pq} &=& F^{I(\B)}_{pq} -\frac{1}{2}\left\{S^{(\B)},F^I\right\}_{pq} \nonumber\\  & &-\frac{1}{2}\sum_{ir}\Big(g_{priq}-g_{pirq}\Big)S^{(\B)}_{ir},
\end{eqnarray}
\begin{eqnarray}\label{eq:fa_b}
    \covd F^A_{pq} &=& F^{A(\B)}_{pq} -\frac{1}{2}\left\{S^{(\B)},F^A\right\}_{pq} \nonumber\\
    & &-\frac{1}{4}\sum_{uv}\sum_r\Big(g_{pruq}-g_{purq}\Big)S^{(\B)}_{vr}\gamma_{uv},
\end{eqnarray}
\begin{eqnarray}\label{eq:q_b}
    \covd Q_{xp} &=& Q^{(\B)}_{xp}-\frac{1}{2}\sum_r S^{(\B)}_{pr}Q_{xr} 
    -\frac{1}{2}\sum_{uvy}\sum_r\Big[(\Gamma_{xuvy}-\Gamma_{xuyv})g_{pury} \nonumber \\ & & -\Gamma_{xvuy}g_{pruy}\Big]S^{(\B)}_{vr}.
\end{eqnarray}
We note that these equations consist of an explicit derivative term and various reorthogonalization quantities involving the differentiated metric.
Eq.~(\ref{eq:grad_ci}) does not include, unlike for real perturbations, the difference between the perturbed reference and inactive energy since the trace of an imaginary operator vanishes. This last fact holds also for the configurational gradient perturbed with respect to nuclear magnetic dipole moments.  

The implementation of the direct matrix-vector product between the magnetic CASSCF Hessian and the imaginary perturbed parameters ($\dv{\lambda}{B_i}$) exploits the same techniques as described in Ref. \citenum{Jensen1986} and \citenum{Nottoli2021}, where all the relevant details can be found. We report in the following the final expressions that need to be evaluated in order to assemble the matrix-vector product between the Hessian and the perturbed parameters
\begin{align}
    &\sum_{r>s}G_{pq,rs}i\dv{\kappa_{pq}}{B_i} = -2i(\Tilde{F}_{pq} + \Tilde{F}_{qp}), \label{eq:mo_mo}\\
    &\sum_I G_{I,pq}i\dv{c_{I}}{B_i} = 2i(\tr F_{pq} + \tr F_{qp}),\label{eq:ci_mo}\\
    &\sum_{p>q} G_{pq,I}i\dv{\kappa_{pq}}{B_i} = 2i\bra{0}\Tilde{\H}\ket{\Phi_I},\label{eq:mo_ci}\\
    &\sum_J G_{I,J}i\dv{c_{I}}{B_i} = 2i\left(\mel{\Phi_I}{\H}{\dv{\mathbf{c}}{B_i}} - E_0\dv{c_{I}}{B_i}\right).\label{eq:ci_ci}
\end{align}
In Eq.~(\ref{eq:mo_mo}), we introduced the one-index transformed Fock matrices:
\begin{equation}
    \Tilde{F}^I_{pq} = \left\{\dv{\kappa}{B_i},F^I\right\}_{pq}  + \sum_{ir}\Big(g_{priq}-g_{pirq}\Big)\dv{\kappa_{ir}}{B_i},
\end{equation}
\begin{equation}
    \Tilde{F}^A_{pq} = \left\{\dv{\kappa}{B_i},F^A\right\}_{pq} + \frac{1}{2}\sum_{uvr}\Big(g_{pruq}-g_{purq}\Big)\gamma_{uv}\dv{\kappa_{vr}}{B_i},
\end{equation}
\begin{eqnarray}
    \Tilde{Q}_{xp} &=& \sum_r Q_{xr}\dv{\kappa_{pr}}{B_i} \nonumber\\
    & &+\sum_{uvy}\sum_r\Big[(\Gamma_{xuvy}-\Gamma_{xuyv})g_{pury} -\Gamma_{xvuy}g_{pruy}\Big]\dv{\kappa_{vr}}{B_i},
\end{eqnarray}
that has a similar structure and symmetries as Eqs.~(\ref{eq:fi_b}) to  (\ref{eq:q_b}).
In Eq.~(\ref{eq:ci_mo}), we have introduced the generalized Fock matrix built with antisymmetrized transition density matrices, that is
\begin{equation}
    \tr F_{pq} = \sum_r \tr \gamma_{pr}h_{qr} + \sum_{rst}\tr \Gamma_{prst}g_{qrst},
\end{equation}
with
\begin{align}
    & \tr \gamma_{uv} = \mel{0}{\E_{uv}}{\dv{\mathbf{c}}{B_i}} - \mel{\dv{\mathbf{c}}{B_i}}{\E_{uv}}{0}, \\
    & \tr \Gamma_{uvxy} = \mel{0}{\hat{e}_{uvxy}}{\dv{\mathbf{c}}{B_i}} - \mel{\dv{\mathbf{c}}{B_i}}{\hat{e}_{uvxy}}{0},
\end{align}
and where $\ket{\dv{\mathbf{c}}{B_i}}=\sum_I\ket{\Phi_I}\dv{c_I}{B_i}$.
Finally, in Eq.~(\ref{eq:mo_ci}), $\Tilde{\H}$ is the one-index transformed Hamiltonian:
\begin{equation}
    \Tilde{\H} = \sum_{uv}\E_{uv}\Tilde{F}^{I}_{uv} + \frac{1}{2}\sum_{uvxy}\hat{e}_{uvxy}\Tilde{g}_{uvxy}.
\end{equation}
The integrals referenced in Eqs.~(\ref{eq:grad_ci}) and (\ref{eq:mo_ci}) show different permutational symmetries compared to the real case. In particular, the one-electron integrals are antisymmetric while the two-electron integrals only exhibit fourfold permutational symmetry. The direct FCI routines used in our CASSCF code have been generalized accordingly. 
We also note that Eqs.~(\ref{eq:mo_mo}) to (\ref{eq:ci_ci}) have been expressed in the determinantal basis -- {\it i.e.}, $I$ and $J$ are running over all Slater determinants. 

\subsection{Implementation of GIAO-CASSCF with Cholesky-decomposed integrals}
The two-step Cholesky decomposition algorithm proposed by Folkestad {\it et al.},\cite{Folkestad2019} and further refined by Zhang {\it et al.}\cite{Zhang2021} has been implemented in the Mainz INTegral (MINT) package\cite{mint} by some of us and extended for the calculation of magnetic integral derivatives.\cite{Burger2021,mint_derb}
The derivative of the two-electron repulsion integral matrix can be written in a form similar to corresponding density-fitting expressions
\begin{equation}\begin{split}\label{eq:res_id}
    \left(\pdv{(\mu\nu|\rho\sigma)}{\B}\right)_{\B=0} =& \sum_{PQ}^{N_{ch}} \left(\pdv{\mu\nu}{\B}\Big|P\right)_{\B=0}M_{PQ}^{-1}\left(Q|\rho\sigma\right) + \\ &+\sum_{PQ}^{N_{ch}} \left(\mu\nu|P\right)M_{PQ}^{-1}\left(Q\Big|\pdv{\rho\sigma}{\B}\right)_{\B=0},
\end{split}\end{equation}
$M^{-1}_{PQ}$ being the inverse metric matrix in the non-orthogonal Cholesky basis. 
From Eq.~(\ref{eq:res_id}), the perturbed and unperturbed Cholesky vectors can be defined respectively as
\begin{align}
    &\left(\pdv{L^P_{\mu\nu}}{\B}\right)_{\B=0} = \sum_Q \left(\pdv{\mu\nu}{\B}\Big|Q\right)_{\B=0}K_{QP}^{-T}, \\
    &L^P_{\mu\nu} = \sum_Q \left(\mu\nu|Q\right)K_{QP}^{-T},
\end{align}
with $\mathbf{K}^{-T}$ -- a shorthand notation for $(\mathbf{K}^{-1})^{T}$ -- as the inverse Cholesky factor of the metric, {\it i.e.}, $\mathbf{M} = \mathbf{K}^T\mathbf{K}$. As shown in Ref.~\citenum{Burger2021} and further demonstrated in Ref.~\citenum{mint_derb}, the perturbed Cholesky vectors constitute antisymmetric matrices and hence only the lower triangular part is stored.

All the equations presented in the previous section have been implemented exploiting Cholesky-decomposed integrals and using the techniques previously described in Ref.~\citenum{Nottoli2021}. In particular, the loop over the Cholesky vectors is kept external in the code and is parallelized by means of shared-memory OpenMP instructions. Furthermore, for each Cholesky vector, efficient level 2 and 3 BLAS routines are used to perform various contractions. As an example, let us illustrate the implementation of the first terms of Eqs.~(\ref{eq:fi_b}), (\ref{eq:fa_b}), and (\ref{eq:q_b}). In particular, $F^{I(\B)}_{pq}$ and $F^{A(\B)}_{pq}$ are built in the atomic-orbital basis.
\begin{eqnarray}\label{eq:fb_i_cd}
    F^{I(\B)}_{\mu\nu} &=& h^{(\B)}_{\mu\nu} + \sum_P\sum_{\rho\sigma}D^{I}_{\rho\sigma}\Bigg[\pdv{L^P_{\mu\nu}}{\B}L^P_{\rho\sigma} - L^P_{\mu\nu}\pdv{L^P_{\rho\sigma}}{\B} \nonumber \\
    & & -\frac{1}{2}\pdv{L^P_{\mu\rho}}{\B}L^P_{\sigma\nu} + \frac{1}{2}L^P_{\mu\rho}\pdv{L^P_{\sigma\nu}}{\B}\Bigg],
\end{eqnarray}
\begin{eqnarray}\label{eq:fb_a_cd}
    F^{A(\B)}_{\mu\nu} &=& \sum_P\sum_{\rho\sigma}D^{A}_{\rho\sigma}\Bigg[\pdv{L^P_{\mu\nu}}{\B}L^P_{\rho\sigma} - L^P_{\mu\nu}\pdv{L^P_{\rho\sigma}}{\B} \nonumber \\
    & & -\frac{1}{4}\pdv{L^P_{\mu\rho}}{\B}L^P_{\sigma\nu} + \frac{1}{4}L^P_{\mu\rho}\pdv{L^P_{\sigma\nu}}{\B}\Bigg],
\end{eqnarray}
with $D^{I}_{\mu\nu}=2\sum_i c^*_{\mu i}c_{\nu i}$ and $D^{A}_{\mu\nu}=\sum_{uv}\gamma_{uv}c^{*}_{\mu u}c_{\nu v}$ being the AO inactive and active density matrices, respectively. The Coulomb contribution is easily computed by first contracting the density matrix with either a perturbed or an unperturbed Cholesky vector. Concerning the exchange term, as usual, it is assembled by first half transforming both the perturbed and unperturbed vectors to the MO basis and then contracting them between each other. Therefore, the formal scaling for the computation of the Coulomb contributions is equal to $\mathcal{O}(N_{ch}N_{\rm b}^2)$. The exchange part of $F^{I(\B)}_{\mu\nu}$ and $F^{A(\B)}_{\mu\nu}$ requires instead $\mathcal{O}(N_{\rm ch}N_{\rm b}^2N_{\rm act} + N_{\rm ch}N_{\rm b}N^2_{\rm act})$ and $\mathcal{O}(N_{\rm ch}N_{\rm b}^2N_{int})$ floating-point operations, respectively. 

The calculation of $Q^{(\B)}_{xp}$ is easily done in the MO basis:
\begin{equation}\label{eq:qb_cd}
    Q^{(\B)}_{xp} = \sum_P\sum_{uvy}\Gamma_{xuvy}\left(L^P_{pu}\pdv{L^P_{vy}}{\B} - \pdv{L^P_{pu}}{\B}L^P_{vy}\right),
\end{equation}
but we need here the perturbed Cholesky vectors in the MO basis with an active index and a generic one. The cost for this transformation is $\mathcal{O}(N_{\rm ch}N^2_{\rm b}N_{\rm act})$, which of course has to be done for the three component of the magnetic field. Using perturbed Cholesky vectors in the MO representation, the computational scaling of $Q^{(\B)}_{xp}$ is $\mathcal{O}(N_{\rm ch}N_{\rm act}^4 + N_{\rm ch}N_{\rm b}N_{\rm act}^2)$. Therefore, the evaluation of Eqs. (\ref{eq:fb_i_cd}) to  (\ref{eq:qb_cd}) shows the same scaling as the construction of the unperturbed CD-based Fock matrices with an additional prefactor of two ({\it c.f.} Eq.~(\ref{eq:res_id})). The active-active block of the perturbed Cholesky vectors in the MO representation is required also to build $\pdv{g_{uvxy}}{\B}$ for the evaluation of Eq.~(\ref{eq:grad_ci}). 

\begin{figure}
    \centering
    \includegraphics[width=0.3\textwidth]{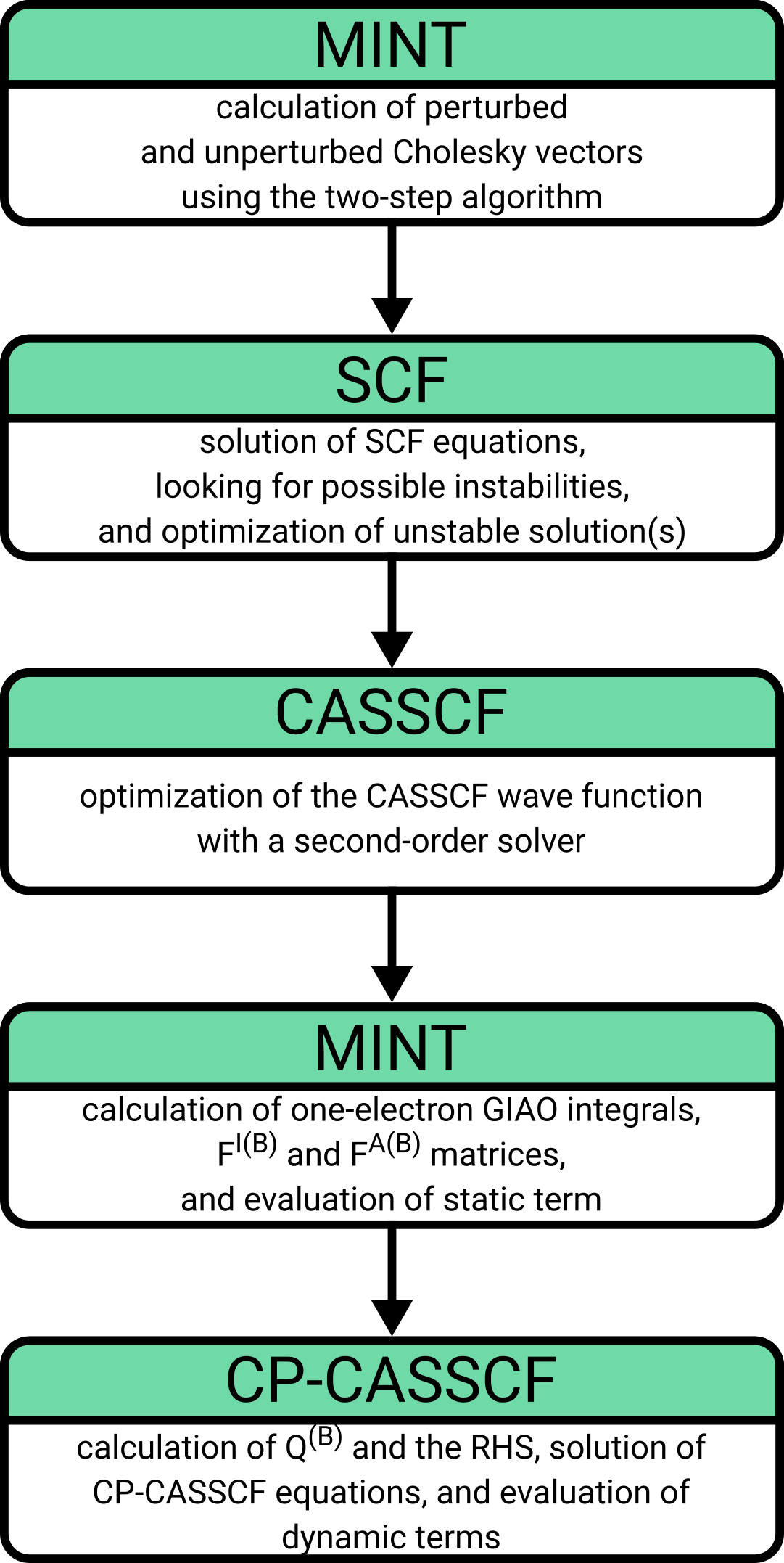}
    \caption{Workflow of a CD-GIAO-CASSCF calculation in {\sc CFour}.}
    \label{fig:workflow}
\end{figure}

The implementation has been carried out in the {\sc CFour} program suite.\cite{Matthews2020,cfour} The workflow for the computation of CASSCF NMR shieldings is the following: First, the MINT program is called where conventional one-electron integrals and both the perturbed and unperturbed Cholesky vectors are calculated and stored on disk. Second, the SCF equations are solved, and if requested, a stability analysis and the computation of Unrestricted Natural Orbitals\cite{Pulay1988,Toth2020} (UNOs) as a guess for the subsequent CASSCF calculation is performed. Third, the CASSCF calculation is done. Fourth, MINT is called again for the calculation of one-electron GIAO integrals, building differentiated Fock matrices, and computing the static contribution to the chemical shieldings. Eventually, the CP-CASSCF program is called for the solution of the coupled-perturbed CASSCF equations and the evaluation of the dynamic terms. The workflow is depicted in Fig.~\ref{fig:workflow} for ease of reference. 

\section{\label{sec:numerical}Numerical Results}

\subsection{Accuracy of GIAO-CASSCF NMR shieldings computed with Cholesky-decomposed integrals}
\begin{figure}
\includegraphics[width=0.3\textwidth]{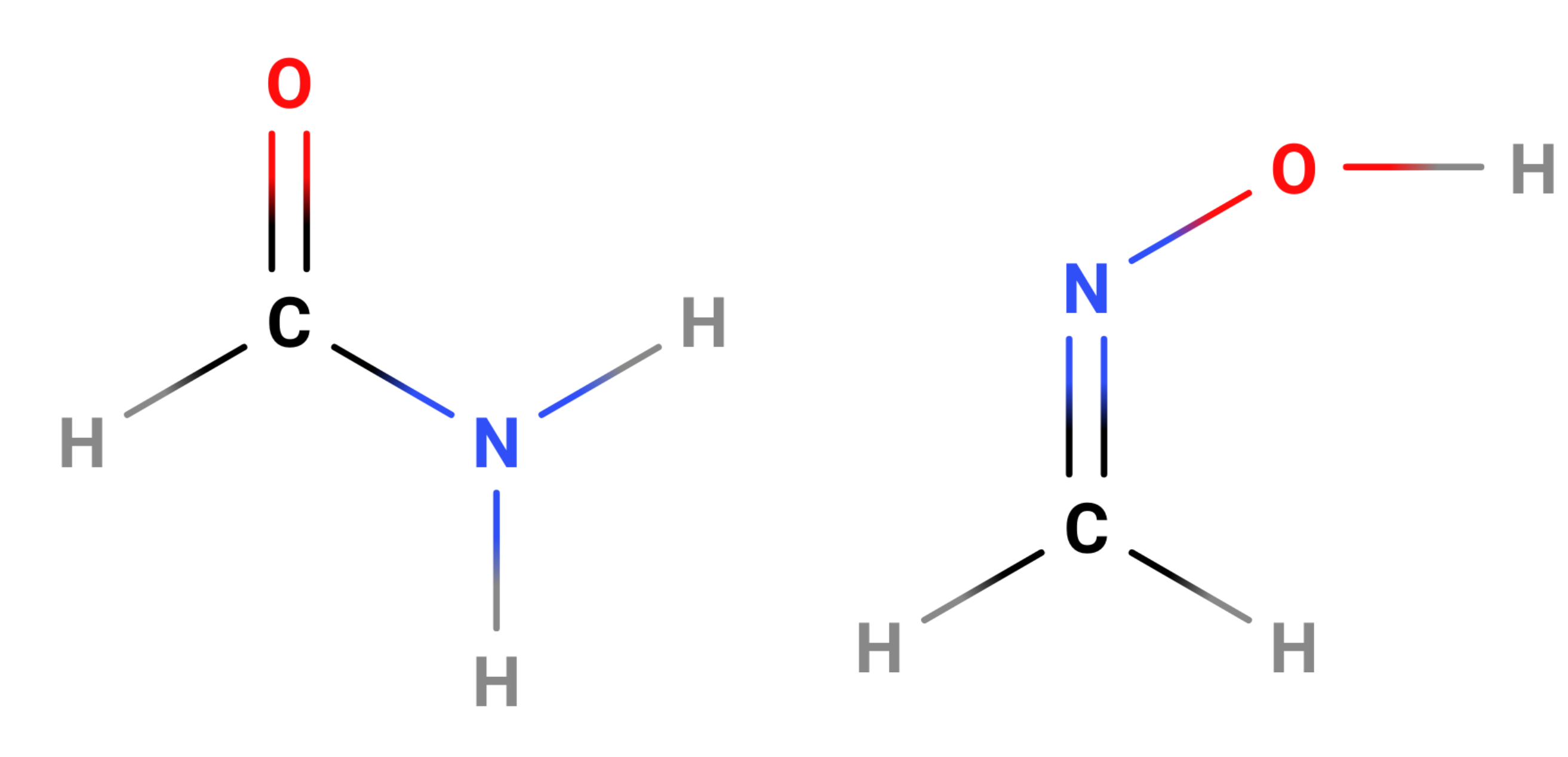}
\caption{\label{fig:isomers} The two molecules used to compare NMR shieldings between an exact calculation and one that uses Cholesky decomposed two-electron integrals. On the left, formamide, on the right formaldoxime.}
\end{figure}

As a first analysis we tested the accuracy of isotropic nuclear magnetic shieldings computed with Cholesky-decomposed integrals. 
The calculations have been carried out on two structural isomers, namely formamide and formaldoxime -- whose molecular representation is shown in Fig.~\ref{fig:isomers}. 
In Table~\ref{tab:cshift_comp}, we report the maximum absolute error for the four different types of nuclei between a reference calculation -- {\it i.e.}, where conventional integrals were used -- and one that exploits the CD with thresholds chosen to be $10^{-4}$ and $10^{-5}$, respectively. To evaluate the effects of the CD approximation with different basis sets, we perform these calculations with increasingly large basis sets from Dunning's hierarchy of correlation-consistent functions.\cite{Dunning1989} The geometries have been optimized at the CASSCF(6,5)/cc-pVDZ level of theory (using conventional integrals) and can be found in the \hyperlink{si}{supporting material} together with the resulting isotropic nuclear magnetic shieldings.

\begin{table*}
\caption{\label{tab:cshift_comp} Maximum absolute error (in ppm) for the isotropic nuclear magnetic shieldings of \isotope[13]{C}, \isotope[17]{O}, \isotope[15]{N}, and \isotope[1]{H} in the two structural isomers formamide and formaldoxime. We report the result of various calculations performed with two different Cholesky-decomposition thresholds ($\delta$) and an increasing number of basis functions.}
\begin{ruledtabular}
\begin{tabular}{ccccccc}
 &\multicolumn{3}{c}{$\delta$ = $10^{-4}$}&\multicolumn{3}{c}{$\delta$=$10^{-5}$}\\
 \cline{2-4}\cline{5-7}
 Nucleus & cc-pVDZ & cc-pVTZ & cc-pVQZ & cc-pVDZ & cc-pVTZ & cc-pVQZ\\ \hline
 \isotope[13]{C} & 0.017 & 0.011 & 0.001 & 0.002 & 0.000 & 0.000\\
 \isotope[17]{O} & 0.074 & 0.009 & 0.009 & 0.005 & 0.002 & 0.000\\
 \isotope[15]{N} & 0.061 & 0.005 & 0.002 & 0.005 & 0.001 & 0.000\\
 \isotope[1]{H}  & 0.001 & 0.003 & 0.001 & 0.000 & 0.000 & 0.000\\
\end{tabular}
\end{ruledtabular}
\end{table*}

The order of magnitude of the errors is consistent with what has been presented by Burger {\it et al.};\cite{Burger2021} {\it i.e.}, using a Cholesky-decomposition threshold equal to $10^{-4}$, the isotropic shieldings exhibit errors of no more than a few hundredths of ppm. When using higher thresholds, as expected, the error decreases. Specifically, with a threshold equal to $10^{-5}$, the errors observed are of the order of $10^{-3}$ ppm. In all cases, the maximum errors are of no relevance for the computational predictions of NMR shieldings. Interestingly, it seems that the error is smaller when using larger basis sets, which is consistent with an observed better agreement between the CASSCF and CD-CASSCF energies in these cases. 

\subsection{Benchmark calculations on small- and medium-sized systems}
\begin{figure}
\includegraphics[width=0.4\textwidth]{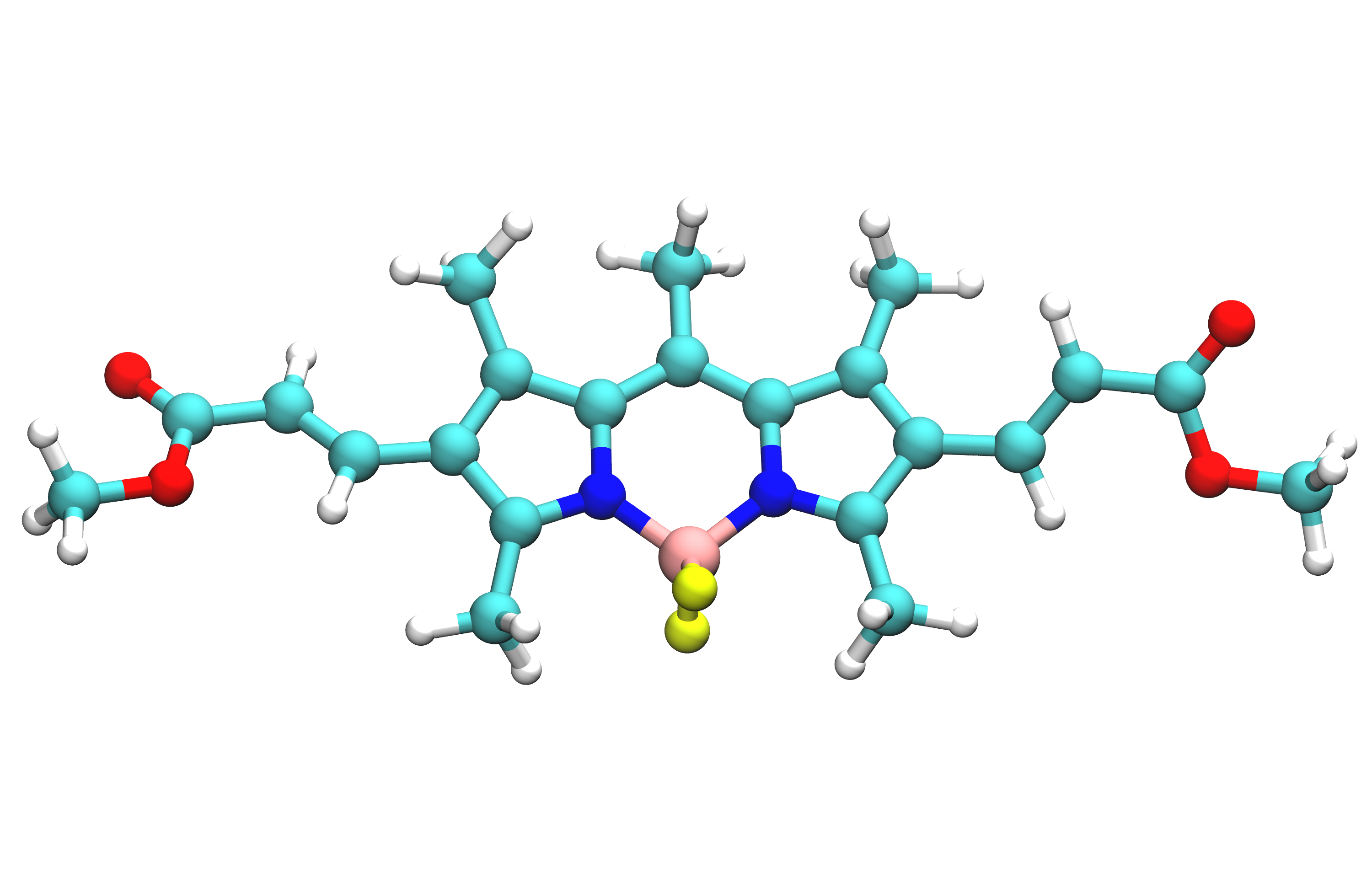}
\caption{\label{fig:bodipy} New BODIPY dye molecule added to the benchmark set. Oxygen in red, carbon in cyan, nitrogen in blue, hydrogen in white, boron in pink, and fluorine in yellow.}
\end{figure}

In order to test the performance of the CD-GIAO-CASSCF algorithm we run benchmark calculations exploiting the same set of molecules already used in Ref.~\citenum{Nottoli2021}. We used a Cholesky-decomposition threshold of $10^{-5}$. The CASSCF equations are considered converged when the root-mean-square (RMS) norm of the configurational and orbital gradient is below $10^{-10}$, while for the CP-CASSCF equations we used a threshold of $10^{-8}$ in the RMS of the residual. These convergence criteria can be considered conservative and are more than sufficient to obtain converged values for the isotropic NMR chemical shifts within the error discussed in the previous subsection. As starting orbitals for the CASSCF calculation we used the UNOs; {\it i.e.}, we look for unstable solution(s) in the RHF wave function, we optimize all the broken-symmetry UHF solution(s), and we get the averaged density matrix from which we can compute the natural orbitals. The results are shown in Table~\ref{tab:bset} and the isotropic nuclear magnetic shieldings are reported in the \hyperlink{si}{supporting material}.  
\begin{table*}
    \centering
    \begin{ruledtabular}
    \begin{tabular}{lcccccccc}
        Molecule       & active space   &  $N_{\rm b}$   &  1$^{\rm st}$ MINT &  SCF\footnote{It includes also the timing required to search instabilities and solve all the UHF unstable solutions. The superscript numbers between parentheses represent the number of instabilities that were found and optimized.} & CASSCF & 2$^{\rm nd}$ MINT & CP-CASSCF & Total  \\ 
        \hline
        adrenaline 	    &6,6 &572 & 52.34 & 1.74$^{(1)}$ & 2.06 & 1.96 & 4.25 & 62.36\\
        anthracene 	    &14,14 &560 &52.58 & 2.80$^{(2)}$ &14.67&1.32 & 27.65 & 99.02\\
        azulene 	    &10,10 &412 &21.97 & 0.45$^{(1)}$ & 0.72 & 0.70 & 1.38 & 25.21\\
        biphenyl 	    &12,12 &500 &37.46 & 0.81$^{(1)}$ & 2.18 & 1.22 & 3.77 & 45.43\\
        catechol 	    &6,6 &324 &11.49 & 0.21$^{(1)}$ & 0.28 & 0.38 & 0.40 & 12.75\\
        dopamine 	    &6,6 &484 &33.14 & 0.79$^{(1)}$ & 1.40 & 1.10 & 1.78 & 	38.2\\
        fluorene 	    &12,12&530&45.72 & 1.32$^{(1)}$ & 2.56 & 2.20 & 3.96 & 55.78\\
        indole 	        &8,8 &368 &16.47 & 0.35$^{(1)}$ & 0.48 & 0.42 & 0.77 & 18.49\\
        l-dopamine 	    &6,6 &574 &44.9 & 2.08$^{(1)}$ & 	3.28 & 2.64 & 4.40 & 57.31\\
        naphthalene 	&10,10 &412 &23.92 & 0.51$^{(1)}$ & 0.84 & 0.73 & 1.24 & 27.25\\
        niacin 	        &6,6 &340 &12.99 & 0.29$^{(1)}$ &0.35 &0.41 & 0.52 & 14.55\\
        niacinamide 	&6,6 &354 &14.47 & 0.33$^{(1)}$ & 0.36 & 0.45 & 0.62 & 16.23\\
        nicotine 	    &6,6 &556 &50.79 & 1.39$^{(1)}$ & 1.79 & 1.95 & 3.28 & 59.20\\
        nor-adrenaline 	&6,6 &514 &39.35 & 1.27$^{(1)}$ & 1.55 & 1.45 & 2.55 & 	46.18\\
        picolinic acid 	&6,6 &340 &12.84 & 0.28$^{(1)}$ & 0.34 & 0.41 & 0.49 & 14.36\\
        pyridine 	    &6,6 &250 &5.22 & 0.10$^{(1)}$ & 0.12 & 0.14 & 0.17 & 5.75\\
        pyridoxal 	    &8,8 &486 &34.09 & 0.88$^{(1)}$ & 1.29 & 1.22 & 2.49 & 39.97\\
        pyridoxamine    &6,6 &528 &42.72 & 1.21$^{(1)}$ & 1.82 & 1.47 & 2.9 & 50.12\\
        pyridoxin 	    &6,6 &514 &39.97 & 1.09$^{(1)}$ & 1.48 & 1.45 & 2.84 & 46.83\\
        resveratrol     &14,14 &678 &74.71 &5.96$^{(2)}$ &19.89&2.89 & 34.79 & 138.24\\
        serotonin 	    &8,8 &558 &48.28 & 1.61$^{(1)}$ & 2.41 & 1.99 & 4.36 & 58.65\\
        tryptophan 	    &8,8 &618 &64.02 & 2.40$^{(1)}$ & 3.58 & 2.38 & 6.87 & 79.24\\
        2Me2HSdiox 	    &4,4 &474 &34.40 & 0.79$^{(1)}$ & 1.20 & 1.15 & 1.80 & 39.35\\
        2Me4HSdiox 	    &6,6 &446 &30.34 &0.64$^{(1)}$& 0.91 &	0.96 & 1.39 & 34.24\\
        coumarin dye    &12,12 &872	&149.86 &13.14$^{(2)}$ &12.35&7.77&21.14 & 204.27\\
        BODIPY dye      & 8,8&1280&386.94& 77.53$^{(3)}$&96.05 &35.57 & 71.96&668.06 \\
    \end{tabular}
    \end{ruledtabular}
    \caption{CD-GIAO-CASSCF computational timings for all the systems in the benchmark set.  
    For each molecule we report the active space, the total number of basis functions ($N_{\rm b}$), and the CPU wall time (in minutes) for the various programs required to compute NMR shieldings at the CD-GIAO-CASSCF level. In particular, the time needed to compute one-electron integrals, perturbed and unperturbed Cholesky vectors (1$^{\rm st}$ MINT); the solution of the SCF equations and the optimization of UHF unstable solution(s) (SCF); the optimization of the CASSCF wave function (CASSCF); the computation of one-electron GIAO integrals, $F^{I(\mathbf{B})}$ and $F^{A(\mathbf{B})}$, and the evaluation of the static term (2$^{\rm nd}$ MINT); the solution of the CP-CASSCF equations and evaluation of the dynamic terms (CP-CASSCF). In the last column, the overall time is also reported. All calculations were carried out on a single cluster node equipped with 4 Intel Xeon Gold 6140 M CPUs running at 2.30 GHz and sharing the work over 28 OpenMP threads.}
    \label{tab:bset}
\end{table*}
At the moment, we have not implemented the out-of-core handling of the (un)perturbed Cholesky vectors. 
A NMR calculation on chlorophyll, one of the benchmark systems presented in Ref.~\citenum{Nottoli2021} is, however, only possible with such an out-of-core implementation.
We therefore replaced chlorophyll with the smaller BODIPY dye, the structure of which is depicted in Fig.~\ref{fig:bodipy}. 
As we did for the other systems in the set, we optimized the new structure at the B3LYP/6-31G(d)\cite{Hehre1972,Becke1993} level of theory using the Gaussian 16 software;\cite{g16} the optimized geometry is given in the \hyperlink{si}{supporting material}.
We report the timings for each step required for the CD-CASSCF calculation of nuclear magnetic shieldings in Table~\ref{tab:bset}. 
Comparing the timings between CASSCF and CP-CASSCF, we notice that the two operations require about the same amount of time. This is to be expected since the two methods exhibit the same scaling and overall computational cost. More specifically, a PCG iteration in CP-CASSCF is about as expensive as a micro-iteration in a norm-extended optimization\cite{Jensen1983} (NEO) second-order optimization step. The ratio between the cost of the two operations depends on the system, and is mainly influenced by the convergence rate of the NEO macro- and micro-iterations on the CASSCF side, which in turn depends on the quality of the starting guess, and on the convergence of the PCG ones in the CP-CASSCF part of the calculation. Moreover, we note that the CP-CASSCF problem is linear, but it has to be solved for the three components of the magnetic field. It is important to remark here the very good performance of the UNO guess, that affords robust and smooth convergence of the nonlinear CASSCF optimization problem. This comes, of course, at a price, as the calculation of the symmetry-broken UHF solutions can be as expensive as the actual CASSCF calculation. Nevertheless, besides the good quality of the guess, using the UNO approach provides additional advantages, namely, a black-box procedure to select the active space, and insights on the multireference character of the system, and in particular on whether a multireference treatment is indeed warranted. Nevertheless, when the UNO procedure becomes too expensive, one could resort to one of the many other alternative approaches to select active spaces that have been proposed in the literature.\cite{Sayfutyarova2017,Stein2016,sayfutyarova2019pi}

A more comprehensive analysis of the computational cost of the overall calculation shows, however, that the real bottleneck in our implementation is neither the CASSCF nor the CP-CASSCF step. In fact, the most expensive step, at the current stage, is the calculation of the unperturbed and perturbed Cholesky vectors, as performed by the first MINT call. With respect to our previous work,\cite{Burger2021} we are using a more efficient implementation of the Cholesky decomposition, which is based on the two-step algorithm proposed by Folkestad et al. \cite{Folkestad2019} and which was extended to the calculation of the perturbed vectors. Looking more in detail into the cost of this step, we observed that it is strongly dominated by the calculation of the two-electron integrals and integral derivatives, the former of which are required in both steps of the CD calculation -- i.e., both to determine the Cholesky basis and to compute the actual Cholesky vectors. 
As discussed in Ref.~\citenum{mint_derb}, there is room for further optimization of the integrals evaluation. Most importantly, this part of the calculation is, at the moment, not parallelized: a parallel evaluation of the shell-quartets required in the procedure would dramatically reduce the overall cost of this step. An optimized, parallel implementation of the evaluation of the Cholesky vectors and their derivatives is being actively pursued. 
The only systems for which the integral evaluation is not completely dominating the calculation are resveratrol, anthracene, and the BODIPY dye, where we observe that the first MINT call takes about 60\% of the overall computer time, to be compared with 80\% or more for the other calculations. Concerning the first two systems, that are described using a larger (14,14) active space, this can be explained by the cost stemming from the configurational part of the CASSCF and CP-CASSCF solvers, which starts to become non-negligible. For the BODIPY dye, on the other hand, the molecular part is dominating the calculation, as it can be seen by the non-negligible impact of the UNO procedure (as well represented by the SCF timing reported in Table~\ref{tab:bset}) as well as of the cost of the CASSCF and CP-CASSCF steps. 

\section{\label{sec:conclusion}Conclusions}
In this contribution we presented the implementation of a CASSCF procedure to compute NMR chemical shifts that uses gauge-including atomic orbitals (GIAO) and the Cholesky decomposition (CD) of the two-electron integral matrix. The implementation closely resembles the one reported in Ref.~\citenum{Nottoli2021} and follows the direct formulation proposed by Jensen {\it et al.}\cite{Jensen1986} The computational cost for the solution of the coupled-perturbed CASSCF (CP-CASSCF) equations is asymptotically equivalent to the cost of a conventional second-order CASSCF optimization algorithm. In particular, when using small active spaces, the expensive operations are the contractions required to build the Fock matrices. We showed that in the overall workflow for the computation of NMR shieldings at the CD-GIAO-CASSCF level, the bottleneck consists in the calculation of the perturbed and unperturbed Cholesky vectors and in particular the evaluation of the two-electron integrals and the differentiated two-electron integrals. For this reason, special effort will be invested in the future in the optimization and parallelization of this part of the code.

At the moment, our implementation relies on the in-core storage of the Cholesky vectors and, for the calculation of the property gradients, both the unperturbed Cholesky vectors and one (out of three) set of perturbed vectors. This is reasonable on standard computer nodes for calculations with up to about 1500 basis functions. Larger calculations would require an out-of-core handling of the Cholesky vectors. We are currently working on an efficient implementation that streamlines and minimizes as much as possible slow disk I/O. 

As a future development, we are actively working on the implementation of geometrical gradients and magnetizabilities, using doubly differentiated Cholesky vectors, both at the CASSCF level of theory and at other correlated, single-reference levels. 

\section*{Supplementary Material}
See \hyperlink{si}{supplementary material} for geometries and NMR isotropic shieldings of the systems used to test the accuracy of CD-GIAO-CASSCF and the performance of the overall algorithm.  

\begin{acknowledgments}
S.S. and J.G. acknowledge funding by the Deutsche Forschungsgemeinschaft (DFG) within project B5 of the TRR 146 (project no. 233 630 050). S.S. also acknowledges support from the DFG via grant STO 1239/1-1.

\end{acknowledgments}

\section*{Data Availability Statement}
The data that support the findings of this study are available within the article and its \hyperlink{si}{supplementary material}.

\appendix

%

\end{document}



\title{Supporting Information: Computation of NMR shieldings at the CASSCF level using gauge-including atomic orbitals and Cholesky decomposition}

\author{Tommaso Nottoli}%
 \email{tommaso.nottoli@phd.unipi.it}
\affiliation{Dipartimento di Chimica e Chimica Industriale, Universit\`{a} di Pisa, Via G. Moruzzi 13, I-56124 Pisa, Italy}

\author{Sophia Burger}
 \email{soburger@uni-mainz.de}
\affiliation{Department Chemie, Johannes Gutenberg-Universit{\"a}t Mainz, Duesbergweg 10-14, D-55128 Mainz, Germany}

\author{Stella Stopkowicz}%
 \email{stella.stopkowicz@uni-mainz.de}
\affiliation{Department Chemie, Johannes Gutenberg-Universit{\"a}t Mainz, Duesbergweg 10-14, D-55128 Mainz, Germany}
\affiliation{Fachrichtung Chemie, Universit{\"a}t des Saarlandes, Campus B2.2, D-66123 Saarbr{\"u}cken, Germany }

\author{J{\"u}rgen Gauss}%
 \email{gauss@uni-mainz.de}
\affiliation{Department Chemie, Johannes Gutenberg-Universit{\"a}t Mainz, Duesbergweg 10-14, D-55128 Mainz, Germany}

\author{Filippo Lipparini}%
 \email{filippo.lipparini@unipi.it}
\affiliation{Dipartimento di Chimica e Chimica Industriale, Universit\`{a} di Pisa, Via G. Moruzzi 13, I-56124 Pisa, Italy}
\
%
\date{\today}

%
%
%
%
%
%
%

\date{\today}

\maketitle

\section{Computational details concerning the accuracy of GIAO-CASSCF NMR shieldings computed with Cholesky-decomposed integrals}

In this section, we report the optimized geometries (in \AA{}ngstr\"om) of the two systems used to test the accuracy of CD-GIAO-CASSCF, namely formamide and formaldoxime, and the isotropic NMR chemical shifts (in ppm) with various Cholesky decomposition threshold and with conventional integrals. Both the geometry optimization and NMR computation is done using Cartesian Gaussians.

\subsection{geometries}

\noindent
a) formamide
\begin{verbatim}
CASSCF/cc-pVDZ 
    C       -0.08605957       0.40912450       0.00000316
    O       -1.13575367      -0.20932040      -0.00000468
    N        1.14518924      -0.16419701       0.00000844
    H       -0.05792672       1.50488505       0.00002554
    H        1.22223946      -1.15926394       0.00000588
    H        1.97395828       0.38648093      -0.00011197
\end{verbatim}

\noindent
a) formaldoxime
\begin{verbatim}
CASSCF/cc-pVDZ 
    C        1.21277966       0.21851388       0.00030556
    H        1.63939419       0.52810284      -0.94290114
    H        1.64980992       0.50343907       0.94642951
    N        0.04967497      -0.58217201      -0.00183535
    O       -1.04808358       0.29951415       0.00119181
    H       -1.78591818      -0.29795429      -0.00058041
\end{verbatim}

\subsection{isotropic NMR shieldings}

\noindent
a) formamide
\\
\noindent
Cholesky threshold = 4
\begin{verbatim}
CASSCF/cc-pVDZ
    C           59.9364        
    O          -60.0644        
    N          197.4375        
    H           24.9324          
    H           28.4821        
    H           28.5919          
\end{verbatim}
\begin{verbatim}
CASSCF/cc-pVTZ
    C           43.3449        
    O          -69.1759        
    N          183.1035        
    H           24.6414          
    H           27.9820         
    H           28.0295          
\end{verbatim}
\begin{verbatim}
CASSCF/cc-pVQZ
    C           39.9222        
    O          -97.6751        
    N          174.1967        
    H           24.4908          
    H           27.3266         
    H           27.4304          
\end{verbatim}

\noindent
Cholesky threshold = 5
 \begin{verbatim}
CASSCF/cc-pVDZ       
    C           59.9350        
    O          -60.0723        
    N          197.4252        
    H           24.9329          
    H           28.4822         
    H           28.5923          
 \end{verbatim}
 \begin{verbatim}
CASSCF/cc-pVTZ       
    C           43.3364        
    O          -69.1714        
    N          183.1083        
    H           24.6412          
    H           27.9807         
    H           28.0270  

 \end{verbatim}
 \begin{verbatim}
CASSCF/cc-pVQZ       
    C           39.9208       
    O          -97.6844        
    N          174.1950       
    H           24.4906         
    H           27.3266         
    H           27.4301    
 \end{verbatim}

\noindent
without Cholesky decomposition
 \begin{verbatim}
CASSCF/cc-pVDZ
    C           59.9352
    O          -60.0763     
    N          197.4233     
    H           24.9328         
    H           28.4823    
    H           28.5925          
\end{verbatim}
 \begin{verbatim}
CASSCF/cc-pVTZ
    C           43.3366        
    O          -69.1704       
    N          183.1091       
    H           24.6412        
    H           27.9805        
    H           28.0271       
\end{verbatim}
\begin{verbatim}
CASSCF/cc-pVQZ
    C           39.9208    
    O          -97.6844     
    N          174.1951       
    H           24.4906     
    H           27.3265     
    H           27.4300     
\end{verbatim}

\noindent
b) formaldoxime
\\
\noindent
Cholesky threshold = 4
\begin{verbatim}
CASSCF/cc-pVDZ
    C           93.9310      
    H           26.5617        
    H           26.5805         
    N         -207.9509        
    O          -31.9781     
    H           19.9756        
\end{verbatim}
\begin{verbatim}
CASSCF/cc-pVTZ
    C           82.1302        
    H           26.6730        
    H           26.6905        
    N         -234.8072        
    O          -45.1546        
    H           19.4145         
\end{verbatim}
\begin{verbatim}
CASSCF/cc-pVQZ
    C           78.4962       
    H           26.6303          
    H           26.6524          
    N         -238.6761     
    O          -46.0696        
    H           19.1629   
\end{verbatim}

\noindent
   Cholesky threshold = 5
 \begin{verbatim}
CASSCF/cc-pVDZ       
    C           93.9161      
    H           26.5620        
    H           26.5808        
    N         -208.0069       
    O          -32.0478        
    H           19.9748    
 \end{verbatim}
 \begin{verbatim}
CASSCF/cc-pVTZ       
    C           82.1195    
    H           26.6703         
    H           26.6912         
    N         -234.8071     
    O          -45.1656      
    H           19.4141       
 \end{verbatim}
 \begin{verbatim}
CASSCF/cc-pVQZ       
    C           78.4965       
    H           26.6304        
    H           26.6525        
    N         -238.6777        
    O          -46.0775      
    H           19.1623       
 \end{verbatim}

\noindent
without Cholesky decomposition
 \begin{verbatim}
CASSCF/cc-pVDZ
    C           93.9139  
    H           26.5620        
    H           26.5808       
    N         -208.0120      
    O          -32.0525      
    H           19.9746        
\end{verbatim}
 \begin{verbatim}
CASSCF/cc-pVTZ
    C           82.1192   
    H           26.6702     
    H           26.6911      
    N         -234.8058      
    O          -45.1633      
    H           19.4142              
\end{verbatim}
\begin{verbatim}
CASSCF/cc-pVQZ
    C           78.4963       
    H           26.6304         
    H           26.6525      
    N         -238.6778      
    O          -46.0772      
    H           19.1622  
\end{verbatim}

\noindent

\section{Computational details concerning the benchmark calculations on small- and medium-sized systems}

In this section, we report Cartesian molecular coordinates (in \AA{}ngstr\"om) for the new BODIPY molecule added to the benchmark set and isotropic NMR shieldings for the whole set of molecules (in ppm). All NMR calculations are performed using spherical Gaussians and a Cholesky decomposition threshold equal to 10$^{-5}$.\\

\subsection{geometry}

BODIPY dye B3LYP/6-31G*
\begin{verbatim}
C       -2.50339726       1.07540102       0.01713342
C       -3.37896601      -0.04852372      -0.05181609
C       -2.57265535      -1.20313123       0.00554098
C       -1.22369869      -0.75225651       0.14753614
N       -1.23890718       0.64901361       0.12874721
C       -0.00637074      -1.44944905       0.29498606
C        1.22332914      -0.75830422       0.28678485
C        2.57907812      -1.21727125       0.27841281
N        1.24669683       0.63956585       0.20345231
C        3.38938403      -0.07079293       0.16403912
C        2.51814372       1.05782631       0.15543525
C       -0.01959324      -2.94763772       0.47165936
H        0.17727335      -3.46819160      -0.47500212
H        0.74625464      -3.26286485       1.17873631
H       -0.97523195      -3.29630171       0.85562159
B        0.00520619       1.58055455       0.23298745
C       -3.07182928      -2.60313918      -0.20877552
H       -2.37460546      -3.20161341      -0.80002868
H       -3.25297057      -3.13938809       0.73337390
H       -4.02218167      -2.57875518      -0.74850357
C        3.09111592      -2.62077708       0.42467341
H        2.70514041      -3.10420020       1.32862001
H        2.81955705      -3.25998578      -0.42479197
H        4.17854584      -2.62126922       0.50560916
C       -2.84668943       2.53019337      -0.00612816
H       -2.39175706       3.03696065       0.85075330
H       -2.43932172       3.00443005      -0.90517252
H       -3.92628365       2.68570524       0.02478917
C        2.87298335       2.50728750       0.07238054
H        2.61292341       2.91079342      -0.91240138
H        2.30443610       3.07843380       0.81178366
H        3.94031909       2.66425330       0.24007214
F        0.04045009       2.43909302      -0.85989966
F       -0.02548584       2.29586674       1.42601155
C        4.83631630       0.05252349       0.07405099
C        5.70475141      -0.80731174      -0.49738924
H        5.26076780       0.96258275       0.49251066
H        5.38483436      -1.70949038      -1.00662796
C       -4.81818109       0.07628325      -0.18948261
C       -5.76366925      -0.82223465       0.16048904
H       -5.17426145       1.01487152      -0.60720781
H       -5.52741572      -1.76417541       0.64154235
C        7.16471554      -0.60247261      -0.54273612
O        7.93586964      -1.37008727      -1.08812028
O        7.56959201       0.53235631       0.08406219
C       -7.21152408      -0.61323622      -0.02251691
O       -8.05743195      -1.41755501       0.32302853
O       -7.51525505       0.57044958      -0.61594726
C        8.98291119       0.77338448       0.05140786
H        9.33573814       0.86516910      -0.97958278
H        9.13326060       1.70866258       0.59205048
H        9.52496151      -0.04227850       0.53778230
C       -8.91586960       0.81503449      -0.80196324
H       -8.98145802       1.79129209      -1.28396430
H       -9.36181014       0.04361138      -1.43582783
H       -9.43758759       0.82590432       0.15907442
\end{verbatim}

\subsection{isotropic NMR shieldings}
\begin{verbatim}
adrenaline, CASSCF(6,6)/cc-pVTZ
C 	 55.2129
C 	 58.5919
C 	 82.5437
C 	 84.8593
C 	 80.1725
C 	 62.2904
H 	 25.1564
H 	 25.3641
H 	 24.6119
O 	 272.8977
H 	 26.7583
O 	 274.7326
H 	 28.2150
C 	 129.5574
C 	 133.7963
O 	 307.2095
H 	 27.6010
N 	 238.3845
H 	 29.5027
H 	 29.5118
H 	 27.7445
C 	 158.8033
H 	 31.1658
H 	 29.1450
H 	 29.6950
H 	 29.2897

anthracene, CASSCF(14,14)/cc-pVTZ
C 	 77.1331
C 	 74.0876
C 	 66.6528
C 	 66.6528
C 	 74.0872
C 	 77.1328
C 	 75.8358
C 	 75.8359
C 	 66.6526
C 	 66.6524
C 	 74.0871
H 	 24.3446
C 	 77.1330
C 	 77.1330
C 	 74.0875
H 	 23.9111
H 	 24.8973
H 	 24.3446
H 	 24.3446
H 	 24.8971
H 	 23.9111
H 	 24.8971
H 	 24.8973
H 	 24.3446

azulene, CASSCF(10,10)/cc-pVTZ
C 	 77.7501
C 	 66.6282
C 	 66.8233
C 	 77.7502
C 	 66.8234
C 	 59.2978
C 	 59.2980
C 	 80.6921
C 	 63.4717
C 	 80.6920
H 	 25.0523
H 	 24.0295
H 	 24.5596
H 	 24.2218
H 	 24.5596
H 	 24.0295
H 	 24.7137
H 	 25.0523

biphenyl, CASSCF(12,12)/cc-pVTZ
C 	 63.8325
C 	 63.8330
C 	 75.3897
C 	 75.3726
C 	 75.3896
C 	 70.6942
H 	 23.9076
C 	 75.3737
C 	 70.6940
H 	 23.9123
C 	 71.7356
H 	 24.6414
C 	 70.6941
H 	 23.9124
C 	 70.6942
H 	 23.9076
C 	 71.7360
H 	 24.6416
H 	 24.6416
H 	 24.7521
H 	 24.7521
H 	 24.6416

catechol, CASSCF(6,6)/cc-pVTZ
C 	 54.3320
C 	 57.4983
C 	 80.5304
C 	 85.1578
C 	 78.2104
C 	 75.6839
H 	 25.0366
H 	 24.8883
H 	 25.3629
H 	 25.1590
O 	 272.1302
H 	 26.7253
O 	 273.2674
H 	 28.1754

dopamine, CASSCF(6,6)/cc-pVTZ
C 	 54.8114
C 	 59.6285
C 	 82.9695
C 	 85.0219
C 	 77.1525
C 	 65.8996
H 	 25.1552
H 	 25.4210
H 	 25.2478
O 	 272.2242
H 	 26.8094
O 	 275.8109
H 	 28.2938
C 	 159.3657
C 	 152.3629
H 	 29.0016
H 	 29.4120
N 	 234.9477
H 	 28.8012
H 	 29.0130
H 	 31.4558
H 	 30.8817

fluorene, CASSCF(12,12)/cc-pVTZ
C 	 160.8140
C 	 59.4483
C 	 58.7036
C 	 58.7036
C 	 59.4481
C 	 76.2140
C 	 74.6384
C 	 74.5194
C 	 80.5182
C 	 80.5182
C 	 74.5195
C 	 74.6383
C 	 76.2138
H 	 28.3925
H 	 28.3925
H 	 24.7611
H 	 24.9412
H 	 24.8707
H 	 24.4288
H 	 24.4288
H 	 24.8707
H 	 24.9411
H 	 24.7611

indole, CASSCF(8,8)/cc-pVTZ
C 	 77.8968
C 	 75.1213
C 	 77.2919
C 	 69.0306
C 	 58.9602
C 	 87.7985
H 	 24.7746
H 	 24.3732
H 	 24.7444
C 	 93.3815
C 	 74.1210
N 	 153.4420
H 	 25.0239
H 	 24.8914
H 	 25.2619
H 	 24.8653

l-dopamine, CASSCF(6,6)/cc-pVTZ
C 	 57.7197
C 	 55.0924
C 	 81.4173
C 	 69.7353
C 	 75.6413
C 	 79.9463
H 	 25.2727
H 	 25.2117
H 	 24.8821
O 	 273.0107
O 	 272.6086
C 	 156.6048
C 	 138.0135
H 	 28.8320
H 	 29.0540
C 	 12.0039
O 	 146.8385
H 	 26.3853
H 	 28.0503
H 	 26.6955
O 	 -112.5139
N 	 232.6858
H 	 29.0404
H 	 30.8709
H 	 31.2245

naphthalene, CASSCF(10,10)/cc-pVTZ
C 	 73.8447
C 	 71.8635
C 	 63.1084
C 	 63.1086
C 	 71.8649
C 	 73.8450
C 	 71.8640
C 	 73.8446
C 	 73.8449
C 	 71.8637
H 	 24.2273
H 	 24.6207
H 	 24.6209
H 	 24.2276
H 	 24.2273
H 	 24.6207
H 	 24.6209
H 	 24.2275

niacin, CASSCF(6,6)/cc-pVTZ
N 	 -80.5633
C 	 46.3376
C 	 48.3884
C 	 76.5476
C 	 62.1275
C 	 73.0657
H 	 22.6651
H 	 24.6906
H 	 23.3895
C 	 14.7956
O 	 159.8741
O 	 -70.0873
H 	 23.2680
H 	 25.2412

niacinamide, CASSCF(6,6)/cc-pVTZ
N 	 -77.8957
C 	 47.4300
C 	 53.2492
C 	 75.9232
C 	 62.1182
C 	 70.6188
H 	 23.3611
H 	 24.6683
H 	 23.1668
C 	 16.2764
N 	 173.6167
O 	 -47.5502
H 	 23.3439
H 	 26.5766
H 	 25.8053

nicotine, CASSCF(6,6)/cc-pVTZ
C 	 75.8010
C 	 50.3567
C 	 66.0260
C 	 62.5311
C 	 49.5113
N 	 -82.4740
H 	 24.8347
H 	 23.5406
H 	 24.1411
H 	 23.6211
C 	 128.0879
N 	 206.0207
C 	 157.5099
H 	 29.2658
C 	 170.4853
H 	 29.8574
H 	 30.2793
C 	 137.9593
H 	 30.1716
H 	 29.9485
H 	 29.8656
H 	 28.8517
C 	 153.0151
H 	 29.6574
H 	 30.3974
H 	 29.4902

nor-adrenaline, CASSCF(6,6)/cc-pVTZ
C 	 55.6677
C 	 59.0183
C 	 84.4959
C 	 84.8451
C 	 75.4731
C 	 63.6989
H 	 25.1776
H 	 25.3482
H 	 24.7207
O 	 272.9861
H 	 26.7891
O 	 274.9096
H 	 28.2217
C 	 127.5868
C 	 147.5311
O 	 295.9132
H 	 27.8977
N 	 242.8555
H 	 31.3423
H 	 30.6458
H 	 29.1315
H 	 28.6648
H 	 27.7608

picolinic acid, CASSCF(6,6)/cc-pVTZ
N 	 -61.3974
C 	 52.4241
C 	 49.4183
C 	 73.0213
C 	 63.3153
C 	 74.2648
H 	 24.5662
H 	 24.1780
H 	 23.4498
H 	 23.4932
C 	 18.2281
O 	 170.8498
O 	 -76.3937
H 	 21.0673

pyridine, CASSCF(6,6)/cc-pVTZ
C 	 76.3174
C 	 49.8757
C 	 65.7571
C 	 76.3168
C 	 49.8619
N 	 -83.5383
H 	 24.8413
H 	 23.4426
H 	 24.4812
H 	 23.4422
H 	 24.8415

pyridoxal, CASSCF(8,8)/cc-pVTZ
C 	 56.6308
C 	 50.1540
C 	 73.3601
C 	 64.7802
C 	 58.3546
N 	 -93.7344
C 	 174.9891
H 	 28.7123
H 	 29.3864
H 	 29.4108
O 	 271.3147
C 	 17.2579
O 	 -387.7920
C 	 132.6871
O 	 315.9320
H 	 27.0120
H 	 27.0295
H 	 21.7802
H 	 23.2636
H 	 27.4608
H 	 30.9891

pyridoxamine, CASSCF(6,6)/cc-pVTZ
C 	 58.0077
C 	 52.7932
C 	 68.3737
C 	 65.4870
C 	 58.9441
N 	 -82.0130
C 	 174.6021
H 	 28.8255
H 	 29.4862
H 	 29.4879
O 	 271.5984
C 	 158.4341
N 	 237.4447
C 	 135.0995
O 	 319.3910
H 	 26.6998
H 	 27.4012
H 	 23.4856
H 	 27.8561
H 	 31.1098
H 	 27.7107
H 	 28.8096
H 	 31.1346
H 	 31.1146

pyridoxin, CASSCF(6,6)/cc-pVTZ
C 	 57.4540
C 	 52.8871
C 	 69.0003
C 	 65.3818
C 	 58.7102
N 	 -85.0383
C 	 174.6484
H 	 28.7924
H 	 29.4626
H 	 29.4693
O 	 271.4612
C 	 140.4967
O 	 303.8271
C 	 134.9031
O 	 318.9985
H 	 26.8475
H 	 27.3224
H 	 23.4398
H 	 27.7622
H 	 31.0757
H 	 26.9707
H 	 27.8938
H 	 30.9514

resveratrol, CASSCF(14,14)/cc-pVTZ
C 	 46.2935
C 	 94.5098
C 	 61.5862
C 	 86.7547
C 	 46.0162
C 	 98.0761
H 	 24.9461
H 	 25.4438
H 	 26.3351
C 	 78.9056
H 	 25.3203
C 	 76.7553
H 	 25.2017
C 	 70.4708
C 	 70.7490
C 	 76.6325
C 	 86.5016
H 	 25.0869
C 	 82.8747
H 	 24.4392
C 	 46.8759
H 	 25.6256
H 	 25.0917
O 	 253.5009
H 	 28.0778
O 	 252.8048
H 	 28.2282
O 	 252.8389
H 	 28.2334

serotonin, CASSCF(8,8)/cc-pVTZ
C 	 50.1770
C 	 85.8000
C 	 90.8815
C 	 67.8735
C 	 62.9582
C 	 88.7291
H 	 25.4059
H 	 24.6046
H 	 24.9645
O 	 257.9605
H 	 28.4005
C 	 82.4969
C 	 77.8417
N 	 160.0834
H 	 25.3428
H 	 25.2405
C 	 165.0810
C 	 152.0953
H 	 28.9740
H 	 29.3650
N 	 234.6506
H 	 28.7403
H 	 28.9300
H 	 31.2874
H 	 30.9057

tryptophan, CASSCF(8,8)/cc-pVTZ
C 	 59.3849
C 	 68.8660
C 	 87.8060
C 	 75.2231
C 	 78.2120
C 	 79.7187
H 	 24.7587
H 	 24.7751
H 	 24.8459
H 	 24.3999
C 	 84.4496
C 	 72.9479
N 	 154.0915
H 	 25.0529
H 	 23.8670
C 	 165.6974
C 	 138.8317
H 	 28.6953
H 	 29.4487
C 	 1.9927
N 	 229.2170
H 	 28.4751
O 	 155.7798
O 	 -81.5444
H 	 25.5490
H 	 30.6820
H 	 30.0683

2Me2HSdiox, CASSCF(4,4)/cc-pVTZ
C 	 60.9917
C 	 84.9859
C 	 60.9921
C 	 84.9856
S 	 472.6692
O 	 277.7383
C 	 129.1443
C 	 129.1444
H 	 28.0003
H 	 28.0503
O 	 277.7402
H 	 28.0503
H 	 28.0003
C 	 182.0041
H 	 29.0818
H 	 29.9229
H 	 29.8726
C 	 182.0046
H 	 29.0817
H 	 29.9228
H 	 29.8728

2Me4HSdiox, CASSCF(6,6)/cc-pVTZ
C 	 64.0838
C 	 65.4501
C 	 64.0839
C 	 65.4502
S 	 504.5738
O 	 280.6666
C 	 129.1796
C 	 129.1796
H 	 27.9483
H 	 27.9267
O 	 280.6662
H 	 27.9267
H 	 27.9483
C 	 98.5952
C 	 98.5951
H 	 27.0373
H 	 26.4552
H 	 27.0373
H 	 26.4552

coumarin dye, CASSCF(12,12)/cc-pVTZ
C 	 49.0583
C 	 98.0076
C 	 43.2039
C 	 87.2441
C 	 72.4650
C 	 89.7179
H 	 25.3723
C 	 64.9476
H 	 24.8642
H 	 25.4859
C 	 80.6379
C 	 14.5040
H 	 24.5361
O 	 93.4297
O 	 -106.1579
N 	 197.5939
C 	 146.6947
H 	 29.1029
H 	 28.3221
C 	 146.3675
H 	 28.3071
H 	 29.1105
C 	 177.4771
H 	 30.7298
H 	 30.9208
H 	 30.2375
C 	 177.1305
H 	 30.9668
H 	 30.6966
H 	 30.3222
C 	 55.7239
C 	 73.0923
S 	 392.5390
C 	 75.1176
H 	 23.7564
C 	 70.6722
H 	 25.5281
H 	 25.1678

BODIPY, CASSCF(8,8)/cc-pVTZ
C 	 32.0827
C 	 72.0471
C 	 38.3738
C 	 67.5872
N 	 60.3590
C 	 29.5072
C 	 67.9438
C 	 38.9214
N 	 59.6748
C 	 72.4050
C 	 32.3799
C 	 175.7863
H 	 29.4870
H 	 28.9914
H 	 28.9868
B 	 116.7351
C 	 176.7220
H 	 29.3790
H 	 29.4207
H 	 29.0444
C 	 177.9562
H 	 29.4914
H 	 29.5371
H 	 28.7663
C 	 182.0277
H 	 28.9809
H 	 29.1255
H 	 29.5174
C 	 181.5441
H 	 29.2685
H 	 28.9407
H 	 29.5578
F 	 368.8113
F 	 370.4667
C 	 68.2464
C 	 80.8005
H 	 24.5293
H 	 26.1912
C 	 68.2341
C 	 81.4796
H 	 24.4832
H 	 26.2334
C 	 16.5096
O 	 -82.3785
O 	 199.4105
C 	 16.3946
O 	 -81.5796
O 	 199.7915
C 	 143.0261
H 	 28.1269
H 	 28.3741
H 	 28.1020
C 	 143.0556
H 	 28.3814
H 	 28.1071
H 	 28.1251
\end{verbatim}